# Experimental Test of Momentum Cooling Model Predictions at COSY and Conclusions for WASA and HESR

H. Stockhorst[*], R. Stassen, R. Maier and D. Prasuhn

*Forschungszentrum Jülich GmbH, Postfach 1913, D-52425 Jülich, Germany*

T. Katayama[1] and L. Thorndahl[2]

*Tokyo[1] and Geneva[2]*

**Abstract.** The High-Energy Storage Ring (HESR) of the future International Facility for Antiproton and Ion Research (FAIR) at GSI in Darmstadt is planned as an anti-proton cooler ring in the momentum range from 1.5 to 15 GeV/c. An important and challenging feature of the new facility is the combination of highly dense phase space cooled beams with internal targets. A detailed numerical and analytical approach to the Fokker-Planck equation for longitudinal filter cooling including the beam - target interaction has been carried out to demonstrate the stochastic cooling capability. To gain confidence in the model predictions a series of experimental stochastic cooling studies with the internal target ANKE at COSY have been carried out. A remarkable agreement between model and experiment was achieved. On this basis longitudinal stochastic cooling simulations were performed to predict the possibilities and limits of cooling when the newly installed WASA Pellet-target is operated.

## INTRODUCTION

The High-Energy Storage Ring (HESR) [1] of the future International Facility for Antiproton and Ion Research (FAIR) at GSI in Darmstadt is planned as an anti-proton cooler ring in the momentum range from 1.5 to 15 GeV/c. An important and challenging feature of the new facility is the combination of phase space cooled beams with internal targets. The required beam parameters and intensities are prepared in two operation modes: the high luminosity mode with beam intensities up to $10^{11}$ anti-protons, and the high resolution mode with $10^{10}$ anti-protons cooled down to a relative momentum spread of only a few $10^{-5}$. Consequently, powerful phase space cooling is needed, taking advantage of high-energy electron cooling and high-bandwidth stochastic cooling. A detailed numerical and analytical approach to the Fokker-Planck equation for longitudinal filter cooling including an internal target has been carried out to demonstrate the stochastic cooling capabilities [2]. The great benefit of the stochastic cooling system is that it can be adjusted in all phase planes independently to achieve the requested beam spot and the high momentum resolution at the internal target within reasonable cooling down times for both HESR modes even in the presence of intra-beam scattering. Also, adjusting the notch filter properly allows to at least partly compensate the mean energy loss due to the internal target. The time of flight (TOF) cooling method is invented to compensate the mean energy loss while filter cooling is applied for the actual momentum cooling. Simulations of this cooling methods are currently under discussion.

[*] email: H.Stockhorst@fz-juelich.de

An elementary introduction to stochastic beam cooling using the beam sampling picture was already presented in [3] where also the main features of the present cooling system of COSY [14] are summarized. This contribution concentrates on a more detailed description of longitudinal beam cooling. The Fokker-Planck equation which is used to model stochastic momentum cooling is explained in a pictorial way.

The COSY-Juelich accelerator has proven to be a dedicated machine to test beam dynamic models, especially in view of the new accelerator HESR at the FAIR facility. The possibility to use internal targets such as cluster targets, gas jet targets and the newly installed Pellet target of the WASA [15] installation makes this machine particularly suitable for stochastic cooling experiments. About ten years of operation of the present stochastic cooling system at COSY together with the internal experiment COSY-11 [16,17] resulted in several improvements of the cooling system, most important the invention of an optical notch filter.

Recently, a series of experimental stochastic cooling studies with the internal ANKE [18] cluster target [4] located in the dispersion free target straight section of COSY to test the model predictions for longitudinal cooling were performed. The routinely operating longitudinal stochastic cooling system applies the optical notch filter method in the frequency band I from 1-1.8 GHz [5].

On the basis of the present model of the beam-target interaction the filter cooling model is employed to predict the longitudinal stochastic cooling capabilities and limits of the present cooling hardware at COSY for the newly installed internal Pellet-target at WASA.

The last chapter gives a brief overview on the stochastic momentum cooling performance in the HESR cooler ring.

## THE LONGITUDINAL STOCHASTIC COOLING MODEL

Longitudinal stochastic cooling can be utilized by two methods [6]. The first method (*Palmer cooling*) uses the fact that the momentum deviation of a particle can be measured directly by a position sensitive pickup located at a point in the ring with high position dispersion. The signal at the output of the pickup averaged over the betatron motion is then proportional to the product $D \cdot \delta$ where $D$ is the dispersion and $\delta$ is the relative momentum deviation of a particle. This correction signal is amplified and sent to the kicker operated in sum mode to provide the necessary momentum or energy correction. In the second method (*Filter cooling*) a pickup in sum mode measures the beam current and the discrimination of particles with different momentum deviations is obtained by inserting a notch filter in the signal path before it drives a kicker in sum mode. The advantage of the filter cooling method, preferred for the HESR design, is that it uses a sum mode pickup which is much more sensitive especially for a smaller number of particles as compared to a pickup that measures the beam position. Moreover, due to filtering after the preamplifier the signal-to-noise ratio is much higher even for a low particle number in the ring. A fact that really helps when the cooling system has to be adjusted for an optimized operation. A further benefit of filter cooling is that the center frequency of the filter can be adjusted to optimize the cooling in the presence of an internal target. A flexibility that is demonstrated below. A disadvantage in filter cooling comes from the notch filter construction. The signal delivered by the pickup is at first equally divided into two paths. One path is delayed by the revolution time corresponding to the nominal beam momentum. Then both signals are subtracted and the resulting signal is amplified and fed to the kicker. Thus a particle sees two correcting kicks at the kicker. The first one when it passes from pickup to kicker and the other one after one turn when it is back at the kicker. Consequently, the undesired mixing from pickup to kicker is larger as compared to the Palmer cooling method where only the undesired mixing on the way from pickup to kicker is relevant. This may lead to a severe restriction in the practical cooling bandwidth when the filter cooling system is applied to a beam with a large initial momentum spread. A fact that is illustrated in more detail below. The filter cooling method can only be practical if the longitudinal Schottky bands are well separated in the cooling bandwidth.

In longitudinal cooling the time evolution of the beam distribution $\Psi(\delta,t)$ is found from (numerically) solving a *Fokker-Planck equation* (*FPE*) [5]

$$\frac{\partial}{\partial t}\Psi(\delta,t) = -\frac{\partial}{\partial \delta}\Phi(\delta,t) \tag{1}$$

with the flux

$$\Phi(\delta,t) = F(\delta)\Psi(\delta,t) - D(\delta,t)\frac{\partial}{\partial \delta}\Psi(\delta,t) \tag{2}$$

where $\delta$ is the relative momentum deviation of a particle. Appropriate initial and boundary conditions are taken into account. The boundary condition describes the finite momentum acceptance of the accelerator.

The flux $\Phi(\delta,t)$ is determined by two terms. The *drift term* $F(\delta)$ describes the *coherent* cooling effect by the self interaction of a single particle with its own momentum deviation. The second term describes the *incoherent* beam heating by diffusion and its strength is determined by the *diffusion coefficient* $D(\delta,t)$ which is always positive. Diffusion always leads to a broadening of the beam distribution.

The FPE, eq. (1) is nothing else but a continuity equation. To understand how the FPE can describe cooling eq. (1) is approximately written as

$$\Psi(\delta, t+\Delta t) \approx \Psi(\delta,t) - \frac{\partial}{\partial \delta}\Phi(\delta,t)\Delta t \qquad (3)$$

to find the change in the particle density within the time internal $\Delta t$. One concludes that the particle density increases for a given momentum deviation in the time internal $\Delta t$ if the flux has a negative slope. In regions where the flux has a positive slope the density is decreased. This is illustrated graphically in figure 1. The simple sketch in figure 1 shows the cooling of an initial Gaussian beam distribution (red curve) when the drift term is proportional to the momentum deviation, $F(\delta) = -k \cdot \delta$, with a positive constant *k*. Using this drift term only the flux as shown in the figure is easily derived graphically. One clearly sees where the beam density is increased or decreased. As a net result cooling occurs as indicated by the blue curve in the left hand side of the figure. A similar sketch can be drawn for the flux if only the (constant) diffusion term is present. One concludes that the diffusion term in $\Phi(x,t) = D\frac{\partial^2}{\partial x^2}\Psi(x,t)$ always leads to a broadening of the beam distribution (sketch the second derivative in figure 1).

From eq. (2) it follows that cooling only occurs if the coherent term predominates the incoherent one, i.e. the resulting flux has a shape similar to that as shown in figure 1.

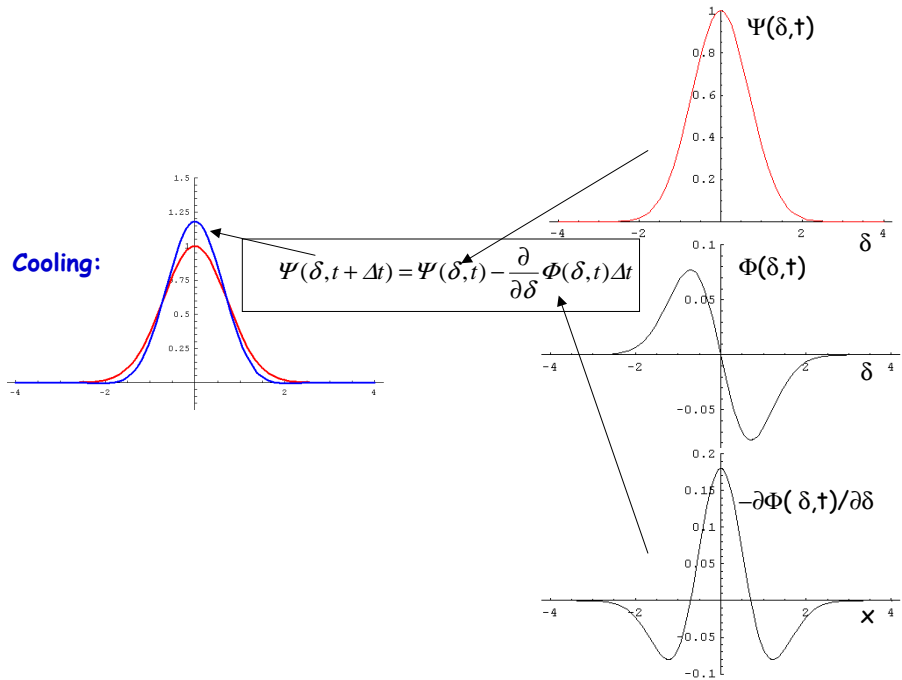

**FIGURE 1.** Sketch of cooling when the drift term is proportional to the momentum deviation. The flux as shown in the middle of the right hand side is simply proportional to the product of the initial beam distribution (red curve) at time t with -δ. Below the flux its derivative is shown. One clearly sees in which regions the density is increased or decreased. If this curve is added to the initial distribution with an appropriate weight (gain) the beam distribution at time t + Δt is found (blue curve). The peak density is increased and the width is reduced.

If drift and diffusion balance each other cooling will stop and an equilibrium is achieved. Note, that reversing the sign in *k* results in a heating of the beam distribution. The sign is determined by the amplifier gain of the cooling system and the frequency slip factor as shown below.

Both drift and diffusion coefficient are determined by the system layout and were calculated in [7] for a specific design of the cooling system at TARN. Later corrections and improvements were given independently by two of the authors (H. St. and T. K.) where it is assumed that pickup and kicker structures are designed as quarterwave loop couplers with electronic transfer functions as given in [8]. The signal path contains a notch filter. Viewed in frequency space this filter exhibits a phase change of 180 degrees in the middle of each Schottky band and the magnitude is symmetric around each revolution harmonic with a sharp drop at the center. The complete theory including the target beam interaction will be published in detail separately [9].

Under the assumption of small momentum deviations and neglecting beam feedback [6] the drift term for cooling of protons or anti-protons is approximately given by

$$F(\delta) \propto -\eta_{tot} \cdot \delta \cdot G_A \cdot S_P \cdot S_K$$
$$\cdot \sum_{n=n_1}^{n_2} n \cos\left(2\theta + n\pi(2r\eta_{PK} + \eta_{tot})\delta + 2\pi n f_0 \Delta T_D\right) \cdot \sin^2(\theta) \tag{4}$$

where the sum runs over all harmonics *n* in the cooling bandwidth $W = (n_2 - n_1)f_0$. The revolution frequency is $f_0$, the ratio distance from pickup to kicker to the ring length is *r* and an additional delay in the signal path is denoted by $\Delta T_D$. The additional delay is used to maximize the cooling effect over the whole cooling bandwidth. The frequency slip factor from pickup to kicker is $\eta_{PK}$ and the total slip factor for the whole ring is $\eta_{tot}$. The momentum and thus frequency dependent phase $\theta$ determines the pickup and kicker magnitude response by $\sin^2(\theta)$.

Each beam harmonic in the bandwidth contributes to cooling and it can be seen that to first order (constant sum) the drift term at each harmonic is proportional to the relative momentum deviation as discussed in figure 1, i.e. the larger the momentum deviation is the larger the correction will be. The correction increases with gain $G_A$ and the product of pickup and kicker sensitivities $S_P \cdot S_K$. The total frequency slip factor $\eta_{tot}$ must be chosen small enough such that there is no band overlap in the whole cooling bandwidth so that each harmonic of the beam is covered by exactly one notch of the filter. As discussed above the product $\eta_{tot} \cdot G_A$ must be positive for cooling for a constant sum. Thus below transition the gain is positive and the sign has to be reversed when operating above transition.

Any deviation from linearity (*undesired mixing effect*) of the drift term results from the contribution of the sum over all harmonics and is determined by both frequency slip factors $\eta_{PK}$ and $\eta_{tot}$. Assuming $\theta \approx \pi/2$ one concludes from eq. (4) that for an additional delay set equal to zero, $\Delta T_D = 0$, the cos-terms at each harmonic are nearly one if the upper frequency of the cooling system is restricted to

$$f_{max} < \frac{f_0}{2 \cdot |2r\eta_{PK} + \eta_{tot}| \cdot |\delta|}. \qquad (5)$$

Note that both frequency slip factors determine this limit due to the fact that two kicks form the correction in notch filter cooling. On contrary, the undesired mixing in Palmer cooling is only determined by $\eta_{PK}$ since this method uses no filter. The result of eq. (5) is also found in reference [10].

If condition eq. (5) is met the sum in eq. (4) is nearly given by $\sum_{n=n_1}^{n_2} n \approx \frac{1}{2}(n_2 - n_1)(n_2 + n_1) \propto f_C \cdot W$ showing the known result that cooling is the better the larger the bandwidth is. The result also shows that one should choose a large center frequency $f_C$, however under the restriction from eq. (5) due to the unwanted mixing.

Two examples illustrate the "mixing dilemma". In figure 2 the drift term (red curve) and the initial distribution (black curve) are displayed for the HESR in the HL-mode with $10^{11}$ anti-protons and kinetic energy $T = 3\ GeV$ [1]. The bandwidth of the cooling system is 2 GHz in the range (2 - 4) GHz and the beam optics is such that $\eta_{PK} = \eta_{tot}$.

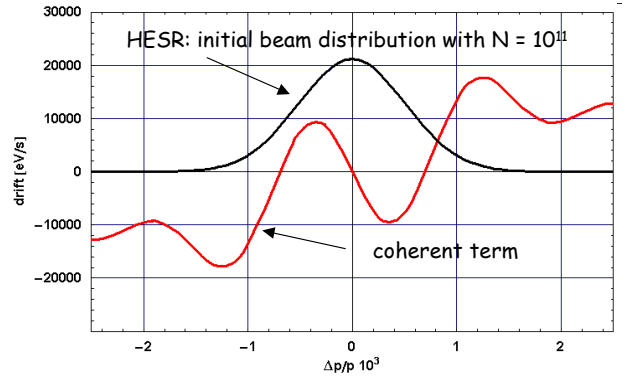

**FIGURE 2.** Drift term (red curve) for the HESR in the HL-mode for a (2-4) GHz cooling system. The black curve is the initial beam distribution at T = 3 GeV. Due to the unwanted mixing cooling can only occur in the restricted range of $\Delta p / p = \pm 7 \cdot 10^{-4}$. Anti-protons with larger momentum deviations are lost at the acceptance limit.

Since the initial beam width is relatively large in the HL-mode at $T = 3\ GeV$ it is clear from the above discussion that cooling can only occur in the range $\Delta p / p = \pm 7 \cdot 10^{-4}$ (cooling acceptance). Anti-protons with larger momentum deviations drift outwards and are lost at the acceptance limit ($\Delta p / p = \pm 2.5 \cdot 10^{-3}$). From eq. (5) it follows that reducing the upper frequency limit of the cooling system can help to increase the cooling acceptance but at the cost of a larger cooling time. The distance from pickup to kicker, i.e. the ratio $r$ in eq. (5), can be reduced. However, the lower limit of the distance between pickup and kicker is fixed due to the finite signal traveling time in the electronics. Another way to increase the cooling acceptance could be a special beam optics with $\eta_{PK} \approx 0$ and $\eta_{tot}$ small. A special cooling procedure to reduce the beam momentum spread with a cooling system having a smaller bandwidth and no filter (TOF cooling) prior to cooling with the filter cooling method is discussed below and seems to be a promising method to solve the mixing dilemma at lower HESR energies for the HL-mode. The mixing dilemma is however much more relaxed at higher HESR energies since the initial momentum spread in the beam reduces with energy.

The main difference between HESR and COSY at the same beam momentum is that the revolution frequency in COSY is about a factor of three larger as a result of COSY's smaller ring size. Therefore, in the case of COSY the condition according to eq. (5) is very well met. Here both frequency slip factors are equal, $\eta_{PK} = \eta_{tot} = 1/\gamma^2 - 1/\gamma_{tr}^2$. With the measured values $f_0 \approx 1.6\ MHz$, $\eta = -0.1$, $\delta = 2\delta_{rms} = 6 \cdot 10^{-4}$ and $r = 0.5$ the upper frequency limit should not exceed about *7 GHz*. In COSY the upper limit of the cooling system is *3 GHz*. This fact results in a drift term as shown in figure 3 which is more or less linear with momentum deviation over the whole beam distribution. Unwanted mixing plays no role in the case of COSY.

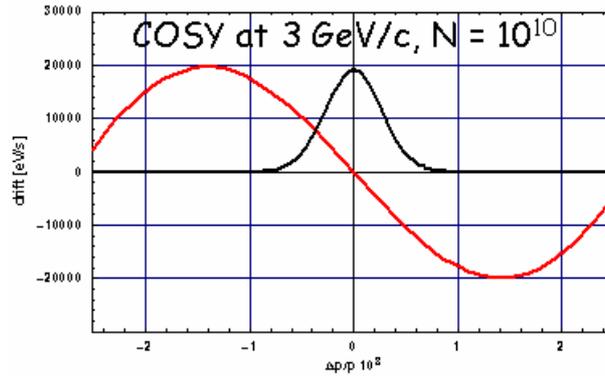

**FIGURE 3.** Drift term (red curve) for COSY for the (1-1.8) GHz longitudinal filter cooling system. The black curve is the initial beam distribution at 3 GeV/c beam momentum and $10^{10}$ stored protons. The drift term is more or less linear over the whole beam distribution.

The diffusion coefficient *D* in the FPE, eq. (1), is similarly a sum over all harmonics and consists of two contributions, Schottky noise heating $D_S$ and thermal noise $D_{th}$ in the electronic system:

Schottky noise: $D_S(\delta,t) \propto G_A^2 \cdot \delta^2 \cdot \Psi(\delta,t)$

Thermal noise: $D_{th}(\delta,t) \propto G_A^2 \cdot T \cdot \delta^2$

Note that both quantities are proportional to the <u>squared amplifier gain</u> and Schottky noise additionally depends on the beam distribution. Obviously thermal noise is proportional to the system temperature *T*. Thus increasing the amplifier gain to increase the coherent term, eq. (4), for more cooling may lead to too much beam heating. As a result the beam is heated instead of being cooled. See also the discussion given in [3].

In the presence of an internal target at a location with zero position and angle dispersion both, the drift term and the diffusion coefficient will be modified. It is assumed that the target beam interaction [11] leads to a mean energy loss per turn $\varepsilon$ resulting in a shift of the whole beam distribution towards lower energies. This effect is taken into account in the FPE as an additional constant drift term $F_T \propto f_0 \cdot \varepsilon$. The longitudinal emittance growth due to energy straggling is given by the mean square relative momentum deviation per target traversal $\delta_{loss}^2$ so that the diffusion term due to the target beam interaction is $D_T = f_0 \cdot \delta_{loss}^2$ for the case of an unbunched beam [11]. Thus diffusion due to the target beam interaction leads to a linear increase of the squared relative momentum spread. Both quantities, mean energy loss per turn and mean square relative momentum deviation per turn can be determined experimentally when cooling is switched off. The measured values are then used in the FPE to describe beam cooling in the presence of the target beam interaction.

In the case of a compensated mean energy loss and a negligible undesired mixing from pickup to kicker as well as $\eta := \eta_{PK} = \eta_{tot}$ one can derive a simple first order differential equation for the relative momentum deviation in the beam. For a beam with $N$ anti-protons or protons this equation is solved for the equilibrium relative momentum spread [9]

$$\delta_{eq,rms} = \frac{4}{5} \left( \frac{3}{16} \cdot \frac{N f_0^2}{|\eta| W f_C} \delta_{loss}^2 \right)^{1/3}. \qquad (6)$$

where the target beam interaction is included by the mean square relative momentum deviation per target traversal $\delta_{loss}^2$.

Note that the equilibrium value does not depend on the initial relative momentum spread. It should be pointed out that if the undesired mixing is significant and/or the mean energy loss is not compensated this formula likely delivers results which are more than a factor of four away from the results found from a solution of the FPE.

## EXPERIMENTAL RESULTS AND MODEL PREDICTIONS

The cooling experiments were carried out at beam momentum *3.2 GeV/c* with about $10^{10}$ stored protons. The frequency slip factor was measured and resulted in $\eta = -0.1$, i.e. the machine was operated above transition, $\gamma > \gamma_{tr}$. Longitudinal cooling was carried out with band I ranging from *1* to *1.8 GHz*. Particle distributions were measured in the frequency range of the harmonic number *1500* with the band II system and can be converted to momentum distributions using the relation $\Delta f / f_0 = \eta \cdot \Delta p / p_0$. The frequency distributions were measured every *2.5 min* or *5 min* in flat top with a duration of about *30 min*.

# Beam Target Interaction

First the target beam interaction was investigated in order to determine the mean energy loss per turn $\varepsilon$ and the mean square relative momentum deviation per turn $\delta_{loss}^2$. The results are shown in the figures 4 and 5. In figure 4 the measured center of the frequency distributions are shown from which the revolution frequency of the protons can derived by dividing the values by the harmonic number 1500. At time zero this gives $f_0 \approx 1.568\,MHz$.

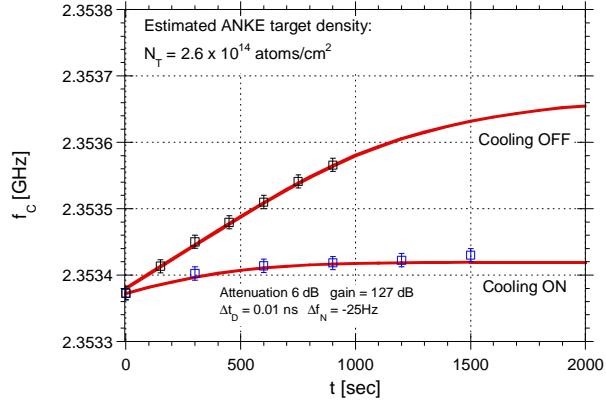

**FIGURE 4.** The measured center frequency at harmonic 1500 (blue symbols: cooling ON, black symbols: cooling OFF) in comparison with the model predictions (red curves).

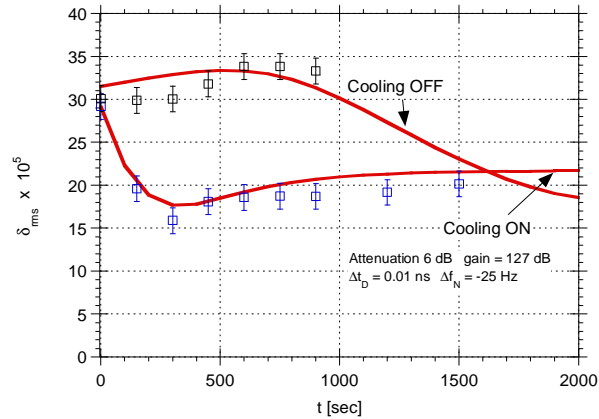

**FIGURE 5.** The measured relative momentum spread at harmonic 1500 (blue symbols: cooling ON, black symbols: cooling OFF) in comparison with the model predictions (red curves). The linear increase of the squared momentum spread determines the mean square relative momentum deviation per turn when cooling is switched off.

The measured data (black symbols) in figure 4 show the expected behavior that the beam distributions are shifted linearly towards lower energies due to the beam target interaction. Note that due to the negative frequency slip factor this corresponds to a linear increase in frequency. Thus the revolution frequency of the protons increases with increasing energy loss. From the slope of the data (black symbols) in figure 4 the mean energy loss per turn was determined to $\varepsilon = -1.8 \cdot 10^{-3}\,eV/turn$. The relative momentum spread in figure 5 (black symbols) shows only a small increase. From the linear increase of $\delta^2_{rms}$ the mean square relative momentum deviation per turn $\delta^2_{loss} = 2 \cdot 10^{-17} / turn$ was derived. The indicated error bars result from three consecutive measurements and reflect the uncertainties due to the finite frequency resolution of the spectrum analyzer. The values for $\varepsilon$ and $\delta^2_{loss}$ have been then used in the FPE when cooling is switched off to determine the beam distributions versus time. A Gaussian initial distribution in the calculations was assumed. The results are shown in figure 4 and 5 as red curves. As can be seen the model deviates from the linear behavior at about *600 s* which is due to particle losses when the shifted distributions reach the momentum acceptance of the machine. This becomes clearly visible when the measured frequency distributions are compared with the distributions predicted by the model as is depicted in figure 6 for two exemplary examples. The measurement as well as the model prediction show a cut off in the distributions at about *2.3537 GHz* which corresponds to the negative relative momentum acceptance limit $\delta_{acc} = -1.4 \cdot 10^{-3}$. It is seen that this value is reached after about *600 s*. Particle losses are increasing then with time as indicated by the increase in the slope at the high frequency side of the distributions. Measured and predicted distributions agree remarkable well.

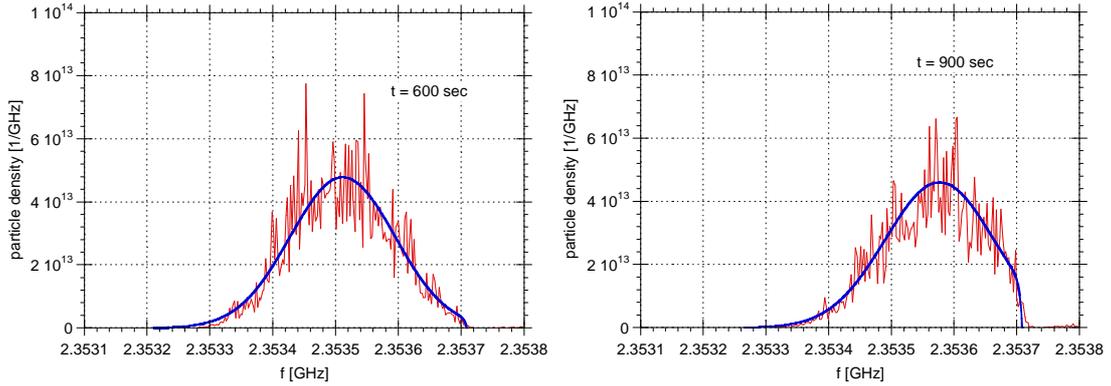

**FIGURE 6.** Measured frequency distributions (red) at harmonic number 1500 for t = 600 s (left) and t = 900 s (right) in comparison with model predictions (blue curves). At t = 600 s the acceptance limit is reached. A sharp cut-off at about 2.3537 GHz is developing as is later clearly visible for t = 900 s.

In figure 6 one also observes that the main effect of the target beam interaction is an increase in the center frequency, i.e. a decrease in the mean momentum of the beam. The beam width only slightly changes during the measuring time.
According to the Bethe-Bloch formula [12] the measured mean energy loss yielded the ANKE target thickness of $N_T \approx 3 \cdot 10^{14} \, atoms/cm^2$.

## Stochastic Cooling with Internal Target

After determining the parameters of the beam target interaction stochastic cooling was switched on. The system delay was adjusted for cooling by means of BTF measurements and the notch filter was set *25 Hz* below the center frequency of the distribution at harmonic one. In momentum space this means that the filter was set above the mean momentum of the protons. Measurements for different attenuations of the electronic gain of the cooling system were then carried out. As an example the following figures show the results for the attenuation set to *6 dB* which corresponds to a model gain and an additional delay of *127 dB* and $\Delta T_D = 0.01 ns$, respectively. Figure 4 shows the center frequency measured at harmonic number *1500* (blue data points) in comparison with the model prediction. The figure clearly shows the cooling effect. The mean energy loss is nearly compensated by cooling. The time development of the relative moment spread during cooling and ANKE target on (blue data points) is fairly well predicted by the model as shown in figure 5. Initially the momentum spread drops down and increases until an equilibrium value $\delta_{rms} = 2.2 \cdot 10^{-4}$ between target beam interaction and cooling is attained after about *1000 s*. Again the cooling effect is clearly visible when the data with cooling on and off are compared. Figure 7 presents a comparison of the measured distribution with the model prediction for different times during cooling.

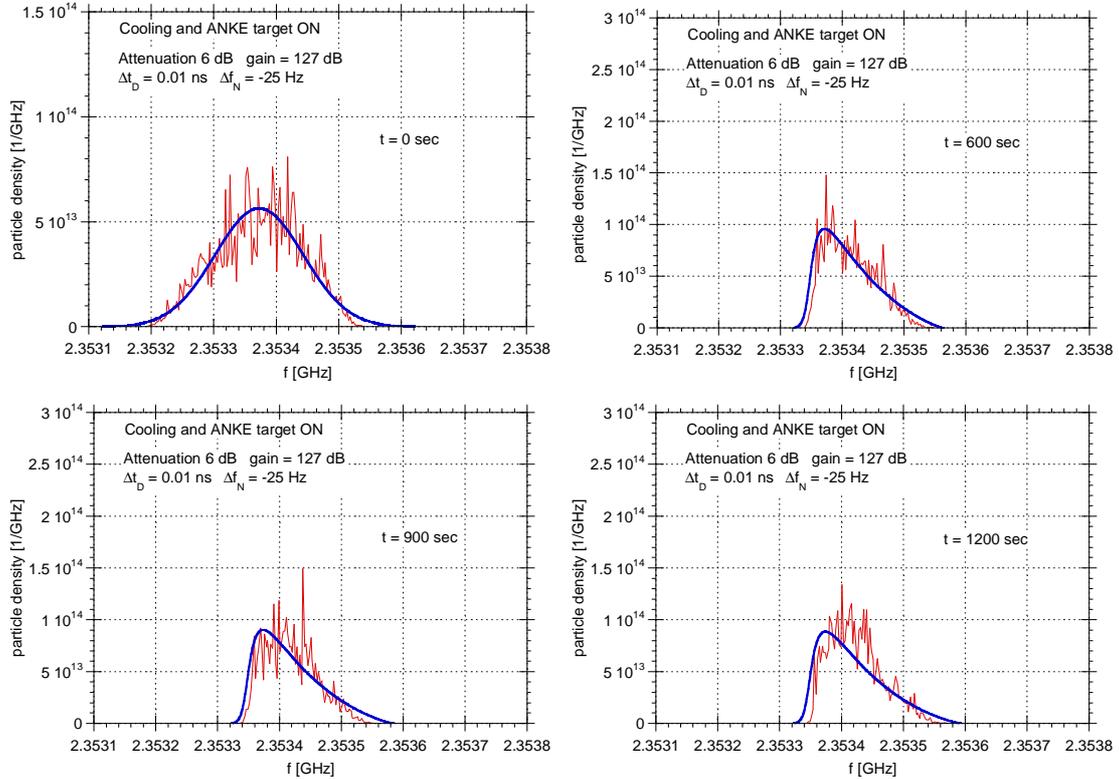

**FIGURE 7.** Measured beam distributions (red) during cooling in comparison with model predictions. Initially a Gaussian distribution has been assumed with values for the center frequency and variance determined from the measured initial distribution. The particle distributions are normalized to the number of protons in the ring. The Fokker-Planck solutions present the absolute beam distributions. There are no scaling factors to adjust the solutions to the measured distributions.

Beam momentum distributions as predicted by the FPE model are shown in figure 8 for different times during cooling as indicated in the figure. Due to the mean energy loss in the target the distributions become asymmetric with respect to zero momentum deviation showing up low momentum tails. However, the mean energy loss is compensated for the measured ANKE target thickness $N_T \approx 3 \cdot 10^{14}\ atoms/cm^2$ and a stable equilibrium momentum spread is attained.

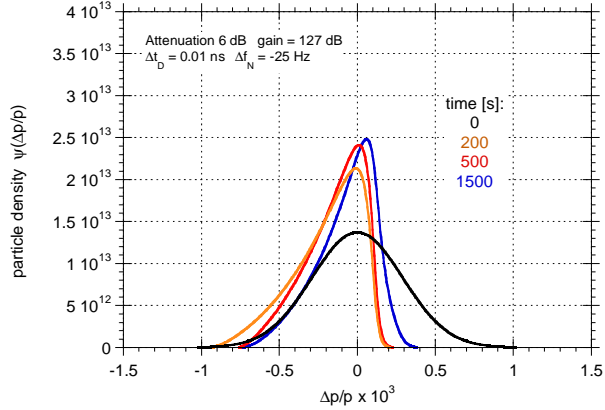

**FIGURE 8.** Beam momentum distributions which correspond to the measured frequency distributions for different times during cooling. Due to the mean energy loss in the target the distributions become asymmetric with respect to zero momentum deviation showing up low momentum tails. The mean energy loss is compensated for the measured ANKE target thickness $N_T \approx 3 \cdot 10^{14}\ atoms/cm^2$ and a stable equilibrium momentum spread is attained.

## MOMENTUM COOLING PREDICTIONS FOR WASA

The new WASA detector and the Pellet target [13] are now successfully installed in COSY and the physics program has already started. Lifetime measurements with the Pellet target switched on and off [13] have shown that a target thickness $N_T \approx 1 \cdot 10^{15}\ atoms/cm^2$ and more can be achieved. As discussed in the previous chapter beam experiments with the ANKE target have shown that the main effect of the target beam interaction is the mean energy loss that leads to a shift of the whole beam distribution towards lower energies. If the mean energy loss is not compensated beam losses will unavoidably occur. Momentum cooling predictions including the target beam interaction due to internal targets with these large thicknesses are therefore of particular interest, not only in view of the WASA experiment at COSY but also in view of the PANDA experiment foreseen at the HESR. First momentum cooling simulations for COSY has been carried out to find the limits of the present cooling system and to show up possible upgrades. It is assumed that as well for a Pellet target the beam-target interaction can be described by the quantities mean energy loss and the mean square relative momentum deviation per target traversal. This assumption can be tested at COSY and is of great importance for model predictions for the HESR.

Figure 9 presents two results using the present momentum cooling system with *(1-1.8) GHz* for the target thickness $N_T \approx 6 \cdot 10^{14}\ atoms/cm^2$ and $N_T \approx 1 \cdot 10^{15}\ atoms/cm^2$. In both cases the number of stored protons is $N = 2 \cdot 10^{10}$ and the beam momentum is $p = 3.2\ GeV/c$. The filter was adjusted to compensate the mean energy loss as much as possible. The model predictions indicate that up to a target thickness of about $N_T \approx 1 \cdot 10^{15}\ atoms/cm^2$ the present cooling system should be capable to compensate the mean energy loss due to the target. The momentum spread in the beam can not be could down but at least the initial relative momentum spread $\delta_{rms} = 3 \cdot 10^{-4}$ can be kept fixed. For larger target thicknesses beam loss will occur.

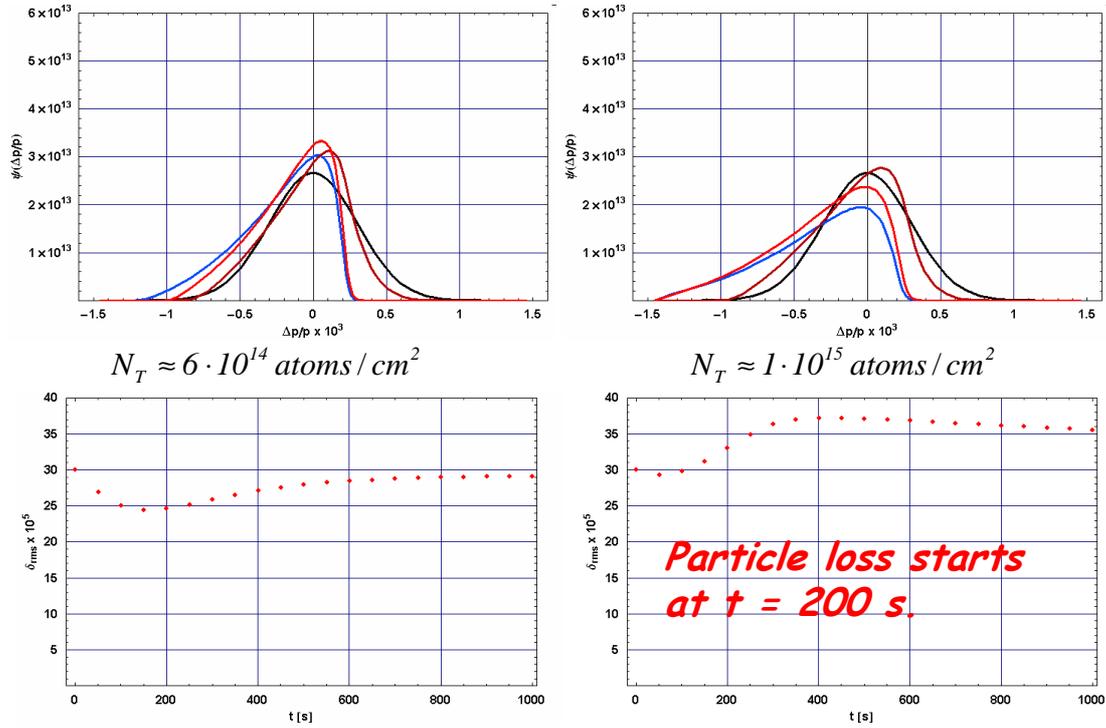

**FIGURE 9.** Momentum cooling simulations for two target thicknesses. Momentum distributions are shown for t = 0 (black), t = 50 s (brown), t = 300 s (red) and t = 1000 s (blue). Using the present cooling system the mean energy loss can be compensated for $N_T \approx 6 \cdot 10^{14}\ atoms/cm^2$. The initial relative beam momentum spread can be kept constant. For a larger target density beam loss can not be avoided.

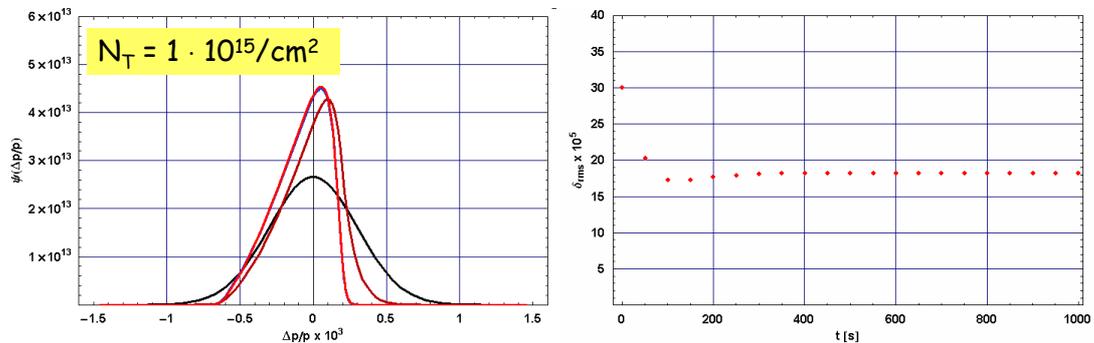

**FIGURE 10.** When operating the longitudinal cooling system at COSY with the full frequency range (1-3) GHz the mean energy loss due to a target thickness of $N_T \approx 1 \cdot 10^{15}\ atoms/cm^2$ can be compensated. The relative momentum spread is reduced to two-thirds of its initial value within 200 s.

A much better performance is found if bandwidth of the cooling system is increased. The simulations in figure 10 show that increasing the bandwidth to its full range of *(1-3) GHz* could help to compensate the mean energy loss even for $N_T \approx 1 \cdot 10^{15} \, atoms/cm^2$ and that the relative momentum spread is reduced to 2/3 of its initial value. A target thickness of $N_T \approx 1 \cdot 10^{15} \, atoms/cm^2$ seems to be the limit for the upgraded stochastic momentum cooling system at COSY. Higher target thicknesses can only be cooled if the mean energy loss is compensated by another method, e.g. using a barrier bucket cavity [13].

Another possibility that seems to provide a promising method to compensate the mean energy loss due to the target beam interaction is the recently discussed time of flight (TOF) cooling method. The notch filter in the cooling chain is removed and is replaced by an 90 degree broadband phase shifter. Simply expressed this method uses the mixing from pickup to kicker to provide the cooling signal. This method prefers a higher bandwidth and a low electronic gain to prevent too much heating of the beam core. A first simulation result is shown in figure 11. It has been assumed that the cooling bandwidth for momentum cooling at COSY has been enlarged to the maximum possible *(1-3) GHz* range.

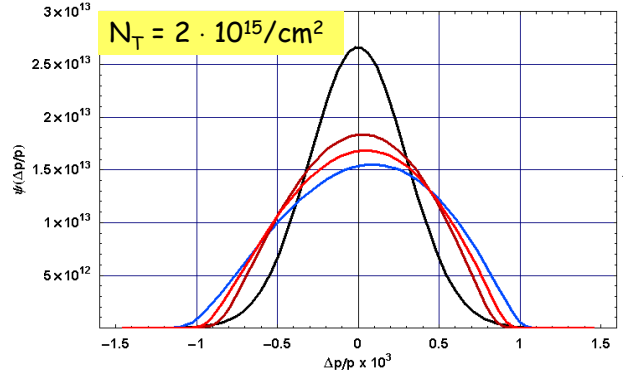

**FIGURE 11.** Momentum distributions for p = 3.2 GeV/c protons during TOF cooling with the momentum cooling system of COSY improved to (1-3) GHz. The distributions are shown at times t = 0 s (black), t = 50 s (brown) t = 100 s (red) and t = 1000 s (blue). Albeit the beam distributions become wider during time the beam center is kept nearly constant even with the large target thickness $N_T = 2 \cdot 10^{15} \, atoms/cm^2$. An equilibrium is reached after about 1000 s.

The figure demonstrates that the mean energy loss is fairly well compensated and a stable equilibrium with almost symmetric distributions is attained after about 1000 s. As compared to the initial distribution only a small increase in the beam momentum spread is observed. It will be investigated at COSY soon in as much these results describe the experimental conditions.

## MOMENTUM COOLING PREDICTIONS FOR HESR

In this chapter a brief summary of the stochastic momentum cooling performance at the HESR is given. The cooler ring HESR [1] will provide anti-protons in the momentum range from *1.5 to 15 GeV/c* for internal target experiments. The required beam parameters and intensities are prepared in two operation modes: the high luminosity mode (HL) with beam intensities up to $10^{11}$ anti-protons, and the high resolution mode (HR) with $10^{10}$ anti-protons cooled down to a relative momentum spread of only a few $10^{-5}$. In both operation modes the target thickness is $N_T = 4 \cdot 10^{15} \, atoms/cm^2$ yielding a luminosity $L = 2 \cdot 10^{32} \, cm^{-2} s^{-1}$ for the HL-mode and $L = 2 \cdot 10^{31} \, cm^{-2} s^{-1}$ for the HR-mode. Stochastic cooling for both transverse and longitudinal phase space will be available for a beam momentum above *3.8 GeV/c* which corresponds to *3 GeV* kinetic energy. Both modes are quite different in its beam quality requirements as seen from the stochastic cooling point of view. While in the HL mode it is only necessary to cool the beam down to some $10^{-4}$ in relative momentum spread the cooling of a high particle number turns out to be a challenge. The anti-protons are delivered from the RESR [1] at a momentum of *3.8 GeV/c* and the beam momentum spread increases with the number of injected anti-protons. In the discussion of the drift term above it has been mentioned that for $N = 10^{11}$ anti-protons the cooling acceptance of the *(2-4) GHz* system is too small to avoid beam losses in the order of *20%*, a number which is not acceptable. Albeit the losses will diminish when the particle energy increases different stochastic cooling scenarios were investigated to avoid particle losses. This was done also in view of a having enough safety margin for the case when the initial beam has a *50%* larger initial momentum spread.

Table 1 summarizes the momentum cooling performance in the HL-mode using only a *(2-4) GHz* system.

**TABLE 1.** HL-Mode Stochastic Momentum Cooling

| Kinetic Energy [GeV] | 3 | 5 | 8 | 12 |
|---|---|---|---|---|
| Band Width [GHz] | 2-4 | 2-4 | 2-4 | 2-4 |
| Particle Number at Equilibrium | $8 \cdot 10^{10}$ | $1 \cdot 10^{11}$ | $1 \cdot 10^{11}$ | $1 \cdot 10^{11}$ |
| $\delta_{rms}$ at Equilibrium | $1.8 \cdot 10^{-4}$ | $1.4 \cdot 10^{-4}$ | $1.3 \cdot 10^{-4}$ | $1.2 \cdot 10^{-4}$ |
| Time to Equilibrium [s] | 500 | 600 | 750 | 800 |

It has been shown that the beam loss at *T = 3 GeV* can be avoided by a pre-cooling system with a (1-2) GHz band and a reduced frequency slip factor.

In the HR-mode the initial momentum spread lies always within the momentum acceptance of the *(2-4) GHz* system. There are no beam losses. A stable equilibrium momentum spread of $\delta_{rms} = 7 \cdot 10^{-5}$ is attained within *200 s* of cooling in the whole energy range above *3 GeV*.

# SUMMARY AND OUTLOOK

The stochastic filter cooling model developed for the investigation of stochastic cooling at the HESR receives a remarkable good agreement with the experimental results at COSY when the internal ANKE target is in operation. The beam target interaction is well described by the model through the quantities mean energy loss and mean square relative momentum deviation per turn. Both quantities can be measured. Once the main parameters are known the model can be employed to predict the cooling properties under different conditions, e.g. if the target thickness is increased, different beam energy, etc.. The good agreement of the model with the experimental results at COSY gives a save confidence that the model will also fairly well predict the cooling properties in the case of the planned HESR at the FAIR facility. However more investigation are needed concerning the undesired mixing that is here much more severe as at COSY. The TOF cooling method will be further investigated in theory as well as in experiment especially including the feedback via the beam. Also other methods to compensate the mean energy loss have to be studied. A further method to compensate the mean energy loss by a barrier bucket cavity will be investigated theoretically and will be soon tested at COSY. Theoretical cooling capability studies with the internal Pellet target of the WASA installation have shown promising results but further investigations are necessary to explore the cooling power also at different energies in COSY. A verification of the cooling model predictions with cooling experiments are needed to gain confidence in the model simulations with an internal Pellet target at WASA similarly to that which is planned to be installed at the HESR.

## ACKNOWLEDGMENTS

One of the authors (H. St.) does not only want to acknowledge the kind invitation of the Jagiellonian University of Krakow and the possibility to give two talks on the COSY11 symposium held in the historic building Collegium Maius, June 17-22, 2007 but also the benign hospitality during the whole stay in the beautiful city of Krakow. Valuable and stimulating discussions with D. Möhl (CERN) and F. Nolden (GSI) are being greatly acknowledged.

## EPILOGUE

It happened that the second talk on the COSY11 symposium was accidentally scheduled by the organizer on June $20^{th}$, 2007. Exactly ten years ago on June $20^{th}$, 1997 the COSY crew was successful in taking into operation the vertical stochastic cooling system. Albeit not optimized and using only the band I *(1-1.8) GHz* system vertical cooling was observed for the first time at the COSY cooler synchrotron. From that on the cooling system was soon fully completed to cool the transverse and longitudinal phase space. The stochastic cooling system became an accelerator tool that routinely could significantly enhance the beam quality for the COSY11 experiment.